\documentclass[aps,showpacs,prl,amsmath,amssymb,amscd]{revtex4}
\usepackage{graphicx}
\newcommand{\lk}{\langle}
\renewcommand{\>}{\rangle}
\renewcommand{\S}{{\bf S}}
\newcommand{\Sd}{{{\bf S}'}}
\newcommand{\Ets}{\overline{E}_{\rm TS}}
\begin{document}
\title{Cancellation of oscillatory behaviors\\ 
 in incommensurate region}
\author{Takahiro Murashima and Kiyohide Nomura}
\affiliation{Department of Physics, Kyushu University,
6-10-1 Hakozaki, Higashi-ku, Fukuoka-city, 812-8581 Japan}
\email{murasima@stat.phys.kyushu-u.ac.jp, 
knomura@stat.phys.kyushu-u.ac.jp}
\date{\today}
\begin{abstract}
In several frustrating systems
incommensurate behaviors are often observed.
For the $S=1$ bilinear-biquadratic model, we show that 
the main oscillatory behavior,
which is proportional to the free edge spins,
is
eliminated in the incommensurate subphase,
considering the average of triplet and singlet energy spectra 
under open boundary conditions.
In the same way,
the $\pi$-mode oscillation is also removed in the commensurate subphase.
Moreover, 
we find that higher order corrections are exponentially decaying
from an analysis of small size data. 
\end{abstract}
\maketitle
\section{Introduction}
Commensurate-incommensurate (C-IC) transitions are interesting phenomena
in frustrating quantum spin systems. 
In the Haldane gap systems,
incommensurabilities have often been regarded as troublesome problems
and have rarely been discussed in detail until recently.
In 
inelastic neutron scattering experiments,
Xu {\it et al}\cite{Xu2000} have revealed that
a quasi-one-dimensional oxide, Y$_{2-x}$Ca$_x$BaNiO$_5$,
has an incommensurate double-peaked structure factor.
However, an analytical interpretation for incommensurabilities
has not been clear.

The
spin-1 bilinear-biquadratic model,
\begin{equation}
{\cal H}=\sum_{i=1}^{N} h_i,\qquad
h_i= \S_i\cdot\S_{i+1}
+\alpha(\S_i\cdot\S_{i+1})^2,
\label{eq_Ham}
\end{equation}
which plays a role of prototype of Y$_{2}$BaNiO$_5$,
 has the C-IC change point which is
 corresponding to the Affleck-Kennedy-Lieb-Tasaki (AKLT) point
 $\alpha=\alpha_{\rm D}=1/3$
\cite{BXG1995,SJG1996}.
The AKLT point is solvable
and has an energy gap above the valence-bond-solid (VBS) 
ground state\cite{AKLT1987,AKLT1988}.
The VBS state is recently getting more attention 
with reference to 
quantum entanglements\cite{VMDC2004,FKR2004,VC2004,FKRHB2006}.

Analyzing the structure factor among the commensurate and incommensurate
 subphases,
and also the C-IC change point, we have deduced two candidates of the ${\it real}$
 structure factor unifying these two regions and the AKLT point
\cite{FS2000,Nomura2003}.
Following the S{\o}rensen-Affleck prescription\cite{SA1994},
we have constructed Green functions from the candidates, and compared them
with the energy gap numerically obtained under the open boundary
 conditions\cite{MN2006}.
Thus, we have found that the Green function consists of two elements
which have anomalies in upper- or lower-half plane. 

In this paper we will show that incommensurate oscillatory behaviors
in the incommensurate subphase
can be canceled, using triplet and singlet energy spectra.
In the same way, we can obtain comparable results in the commensurate
case.
Moreover we discuss higher order corrections considering small size systems.

\section{Short review on singlet-triplet energy gap}

The gapped Haldane phase has nonvanishing nonlocal string order\cite{dNR1989}
and effectively free $S=1/2$ spins at the ends of open chains
\cite{Kennedy1990,HKAHR1990}.
These edge spins bring a low-lying excitation, triplet,
which is degenerate with
the ground state in the thermodynamic limit\cite{Kennedy1990}.
While this excitation is clearly a boundary effect,
it is linked to the bulk behavior because of the existence of nonlocal
string order\cite{SJG1996}.

According to S{\o}rensen and Affleck (SA)\cite{SA1994},
the effective Hamiltonian can be connected with
the Green function:
\begin{equation}
{\cal H}_{\rm eff} = (-1)^N \Sd_{\rm L}\cdot\Sd_{\rm R}
\lambda^2 \int \frac{dq d\kappa}{(2\pi)^2}G(q, \kappa) 
\exp(i q N) \delta(\kappa), \label{eq_Heff}
\end{equation}
where $\Sd_{\rm L}$ and $\Sd_{\rm R}$ are spin-1/2 operators at the ends
of chain.
Considering the Clebsch-Gordan coefficient,
we obtain singlet and triplet expectation values for edge spins as
\begin{equation}
\begin{split}
\lk {\rm S}| \Sd_{\rm L}\cdot\Sd_{\rm R}|{\rm S}\>=-3/4, \\
\lk {\rm T}| \Sd_{\rm L}\cdot\Sd_{\rm R}|{\rm T}\>=1/4,
\label{eq_CG}
\end{split}
\end{equation}
where $|{\rm T}\>=|s^{\rm T}=1, s_z=\pm 1,0\>$ and
$|{\rm S}\> = |s^{\rm T}=0, s_z=0\>$\cite{memo1}.
Thus the energy difference between the singlet and triplet states
can be described with the Green function:
\begin{equation}
\begin{split}
\Delta E_{\rm ST} (N) &\equiv E_{\rm T} - E_{\rm S}\\
&=(-1)^N 
\lambda^2 \int \frac{dq}{2\pi}G(q, 0) 
\exp(i q N),
\end{split}
\end{equation}
where $E_{\rm T}$ and $E_{\rm S}$ are the triplet and singlet energies,
respectively.

\begin{figure}[h]
\begin{center}
 \includegraphics[width=8cm]{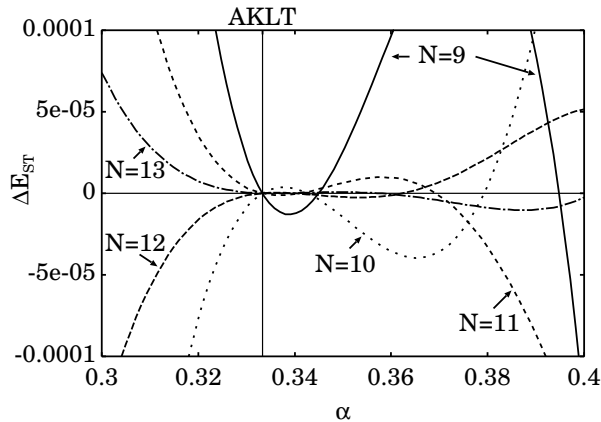}%
\end{center}
 \caption{\label{fig_gap}
Oscillation of the energy gap of edge states.
The energy gaps with different chain lengths 
($N=9, \cdots, 13$)
are plotted as a function of $\alpha$.
}
\end{figure}

Figure \ref{fig_gap} shows the energy gap behaviors 
of the model \eqref{eq_Ham}
with different chain lengths.
Increasing the chain length on some fixed $\alpha$ in the commensurate subphase,
we see that the energy gap oscillates between even and odd chains
and decreases exponentially fast,
while this behavior is not simple in the incommensurate subphase.
The even-odd oscillation is modulated by the frustration
in the incommensurate subphase.

In our previous study\cite{MN2006},
which has been performed to explain these behaviors,
we have found 
\begin{equation}
\Delta E_{\rm ST}
 (N)=
\begin{cases}
(-1)^{N}\widetilde{A}\exp(-\widetilde{m}N)\sin(\sqrt{d}N), 
&\quad (\alpha > 1/3)\\
0, & \quad (\alpha=1/3) \\
(-1)^{N}\widetilde{A}\exp(-\widetilde{m}N)\sinh(\sqrt{d}N). 
&\quad (\alpha < 1/3)\\
\end{cases}
\label{eq_gap}
\end{equation}
The parameters, $\widetilde{A}$, $\widetilde{m}$, and $d$,
depend on $\alpha - \alpha_{\rm D}$ and
have been determined with the nonlinear least-squares fitting method
as shown in Ref. \cite{MN2006}.
We have found that the C-IC change is characterized by the following
Green function:
\begin{equation}
G(q,0) \sim \frac{1}{(q - i\widetilde{m})^2 -d} 
+ \frac{1}{(q + i\widetilde{m})^2 -d}.
\end{equation}

\section{Average of triplet-singlet energies}

Now we define the average of triplet-singlet energies:
\begin{equation}
\Ets \equiv (E_{\rm S}+3E_{\rm T})/4.
\end{equation}
From Eqs. \eqref{eq_Heff} and \eqref{eq_CG}, 
we expect that 
we can eliminate the principal term caused by 
the anomalies of the Green function
using this average.

 \begin{figure}[h]
\begin{center}
 \includegraphics[width=8cm]{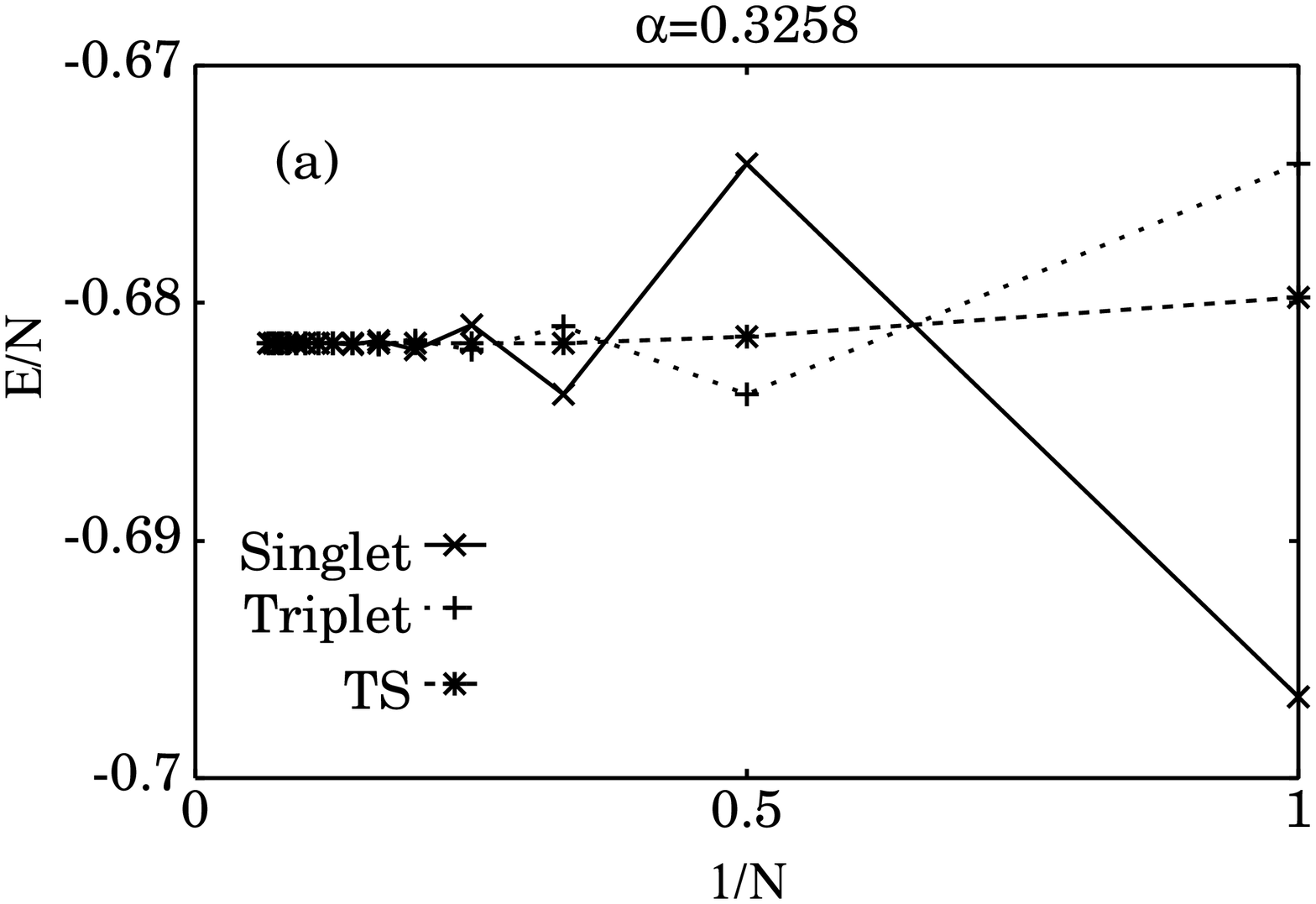}%
 \includegraphics[width=8cm]{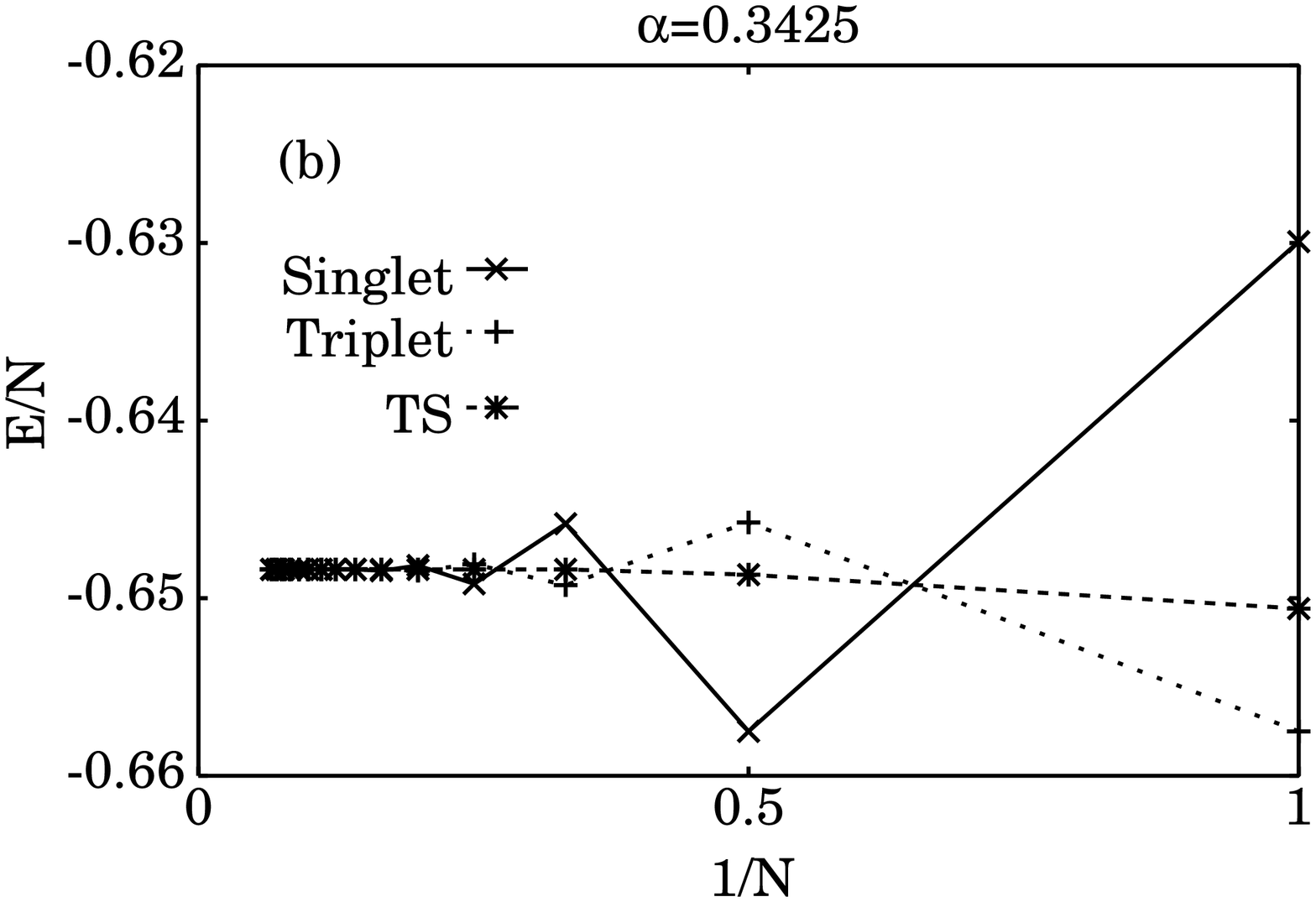}%
\end{center}
 \caption{\label{fig_Ets}
Cancellation of triplet-singlet energies.
Triplet and singlet energy spectra,
and an average of triplet-singlet energies are plotted as a function of
  $1/N$ for $1\le N \le 15$
when (a) $\alpha=$ 0.3258 (commensurate case), and (b) 0.3425 (incommensurate
  case). }
 \end{figure}
Figure \ref{fig_Ets} shows bare triplet and singlet energies and 
averages of triplet-singlet energies 
plotted
as a function of $1/N$ for (a) $\alpha=$ 0.3258 (commensurate case), 
and (b) 0.3425 (incommensurate case).
We see that the average of triplet-singlet energies 
on a fixed $\alpha$
varies linearly with $1/N$ not only in the incommensurate subphase
but also in the commensurate subphase.

\begin{figure}
\begin{center}
 \includegraphics[width=8cm]{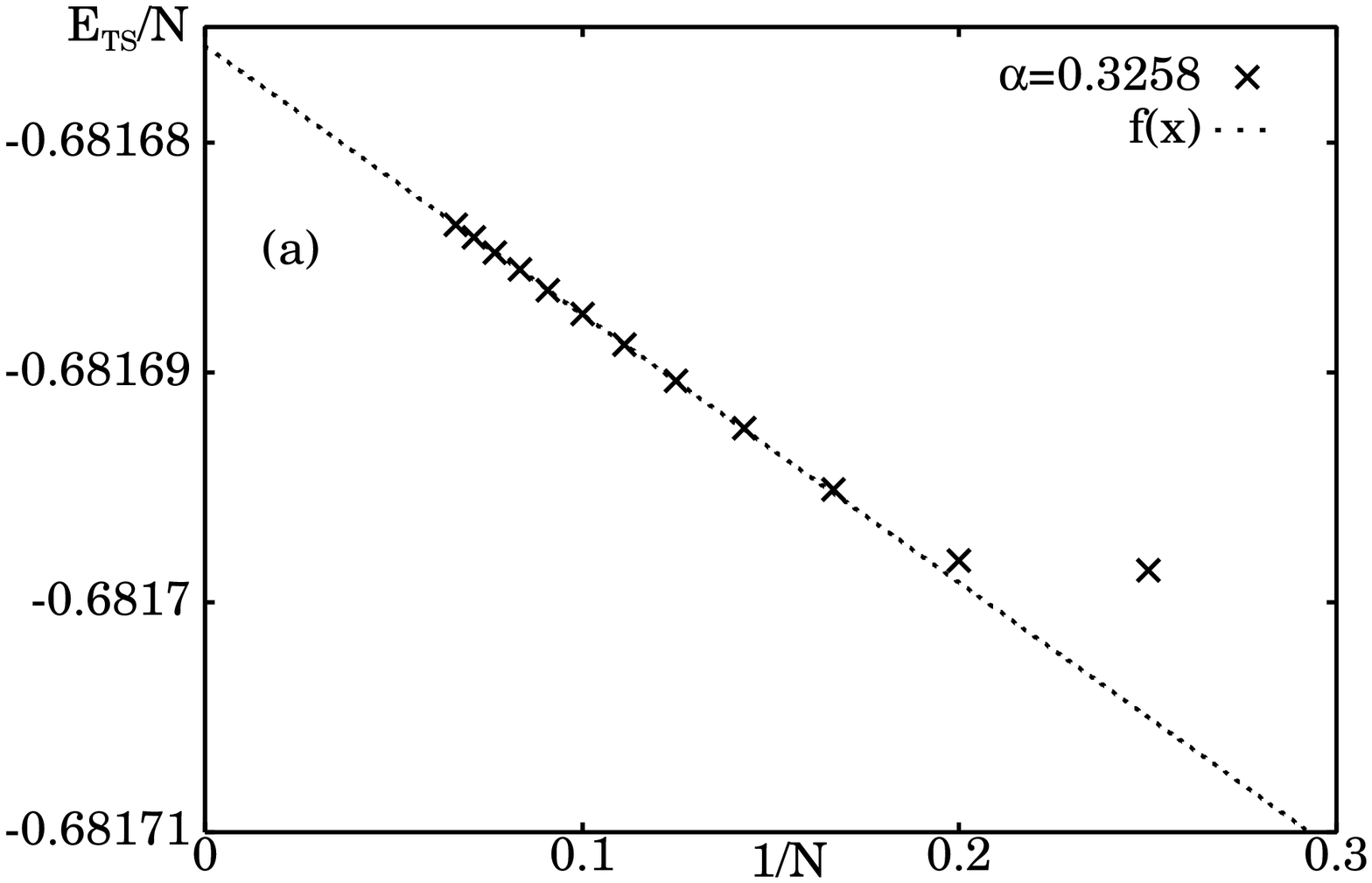}%
 \includegraphics[width=8cm]{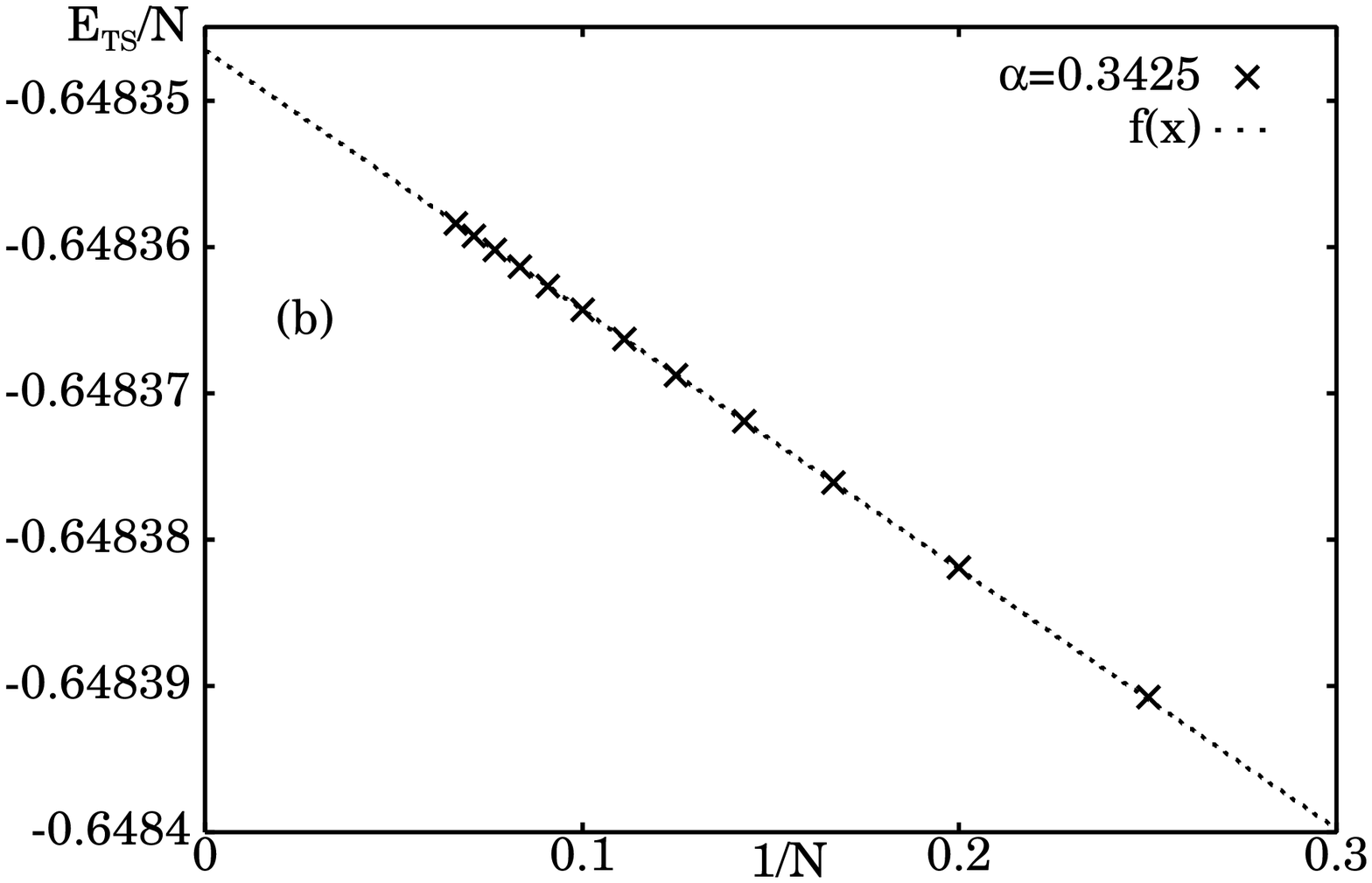}%
\end{center}
 \caption{\label{fig_fitEts}
Finite size effect in an average of triplet-singlet energies
and a least-squares fitting line
  ($c_0+ c_1/N$) are plotted when (a) $\alpha=0.3258$, (b) $0.3425$.
}
\end{figure}
We perform the least-squares fit to the averages of triplet-singlet 
energies using the following fitting function:
\begin{equation}
f(N) = c_0 + c_1 /N. \label{eq_fit}
\end{equation}
Figure \ref{fig_fitEts} shows the data of triplet-singlet averages
and $f(N)$.
When (a) $\alpha=0.3258$,
each coefficients are obtained as
$c_0=-0.68167581   \pm 2 \times 10^{-9}$ and
$c_1=-1.1671 \times 10^{-4} \pm 2 \times 10^{-8}$
for $8 \leq N \leq 15$.
In the same way,
$c_0=-0.6483465831 \pm 2 \times 10^{-10}$ and
$c_1=-1.77295 \times 10^{-4} \pm 2 \times 10^{-9}$ when (b) $\alpha=0.3425$. 
The average of triplet-singlet energies $\Ets/N$ 
seems to behave high linearly in small $\alpha$ region.
In fact, we observe that coefficients 
of $O(1/N^2)$, $O(1/N^3)$ are very small.
Hence,
we can say that a higher power of $1/N$ does not appear.

\begin{figure}[h]
\begin{center}
 \includegraphics[width=8cm]{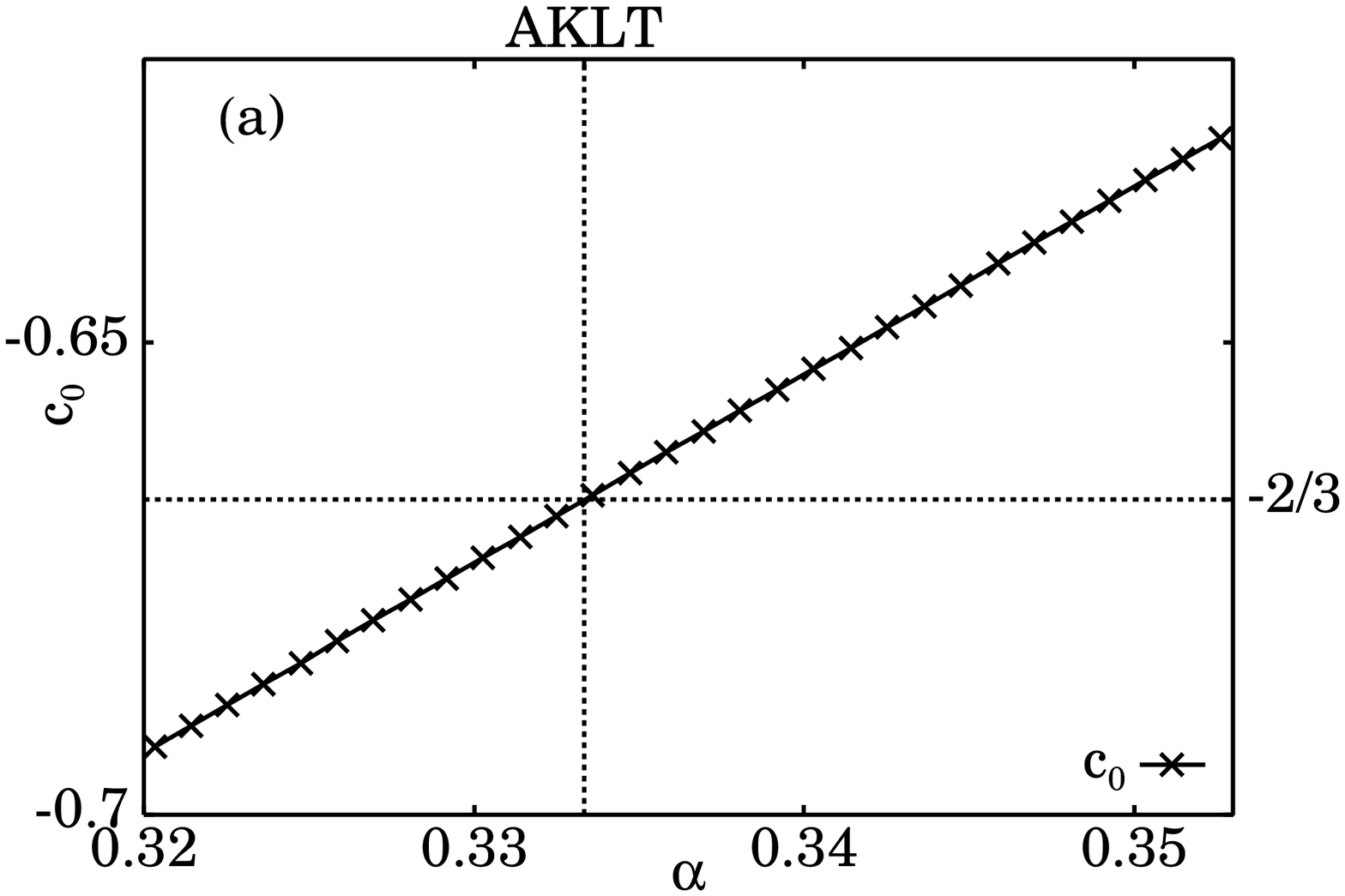}%
 \includegraphics[width=8cm]{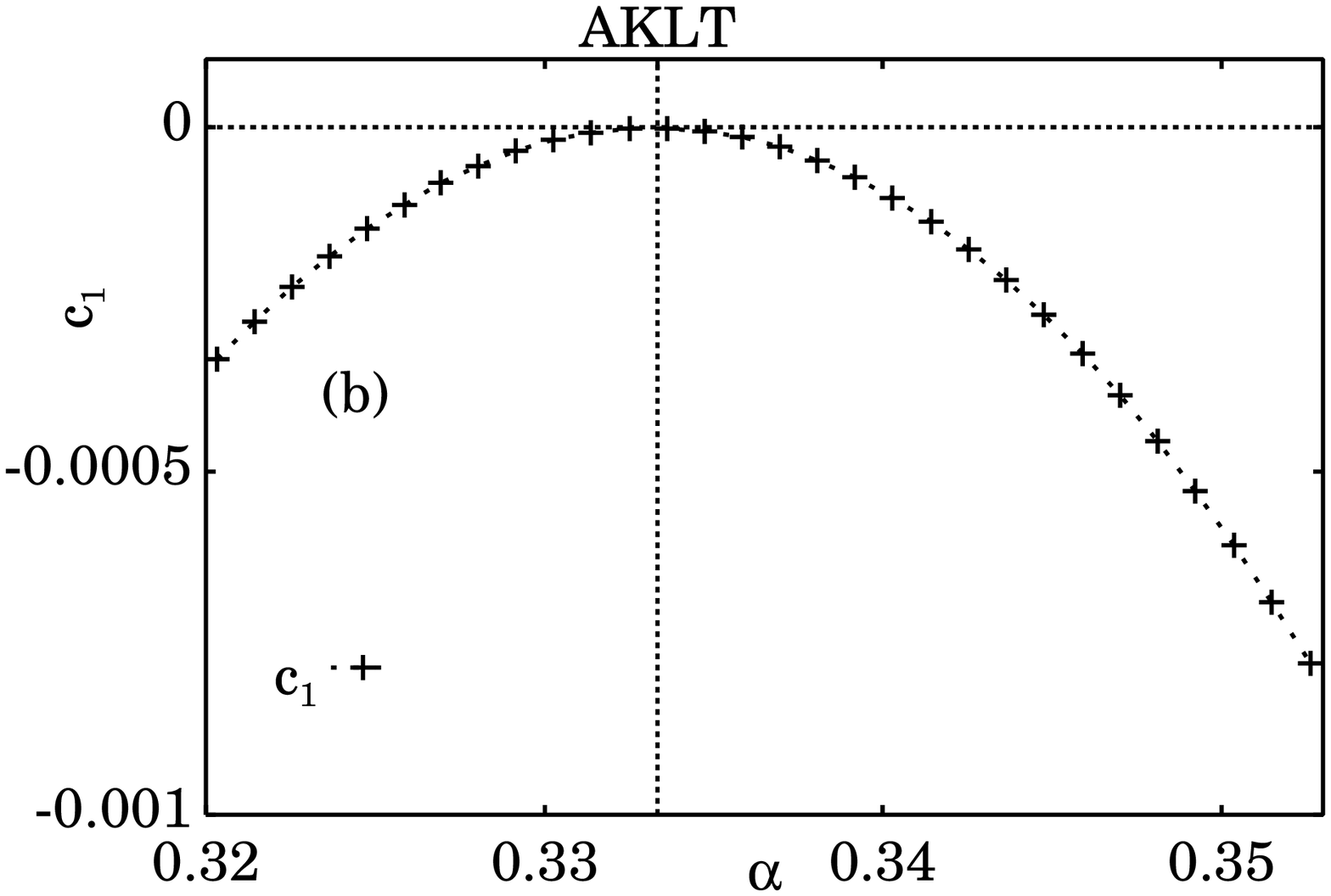}%
\end{center}
 \caption{\label{fig_fit_c}
Coefficients (a) $c_0$ and (b) $c_1$ in Eq. \eqref{eq_fit}
are plotted near the AKLT point. 
It appears that $c_0 \propto \alpha -\alpha_{\rm D}$ and $c_1 \propto 
(\alpha -\alpha_{\rm D})^2$.}
\end{figure}
Figure \ref{fig_fit_c} shows coefficients $c_0$ and $c_1$ in Eq. \eqref{eq_fit}
obtained with the least-squares fitting.
The second term in Eq. \eqref{eq_fit} comes from surface effects.
Since the surface effects are caused by the one-dimensionality,
they are different from the effect of edge spins.
Additionally, the surface effects are perfectly zero,
namely $c_1=0$,
at the AKLT point\cite{memo2}.
Moreover, $c_1$ always shows negative value except the AKLT point.
We see that the fitting parameters
$c_0$ and $c_1$ behave approximately as $\alpha - \alpha_{\rm D}$ and 
$(\alpha-\alpha_{\rm D})^2$, respectively.

\begin{figure}[h]
\begin{center}
\includegraphics[width=8cm]{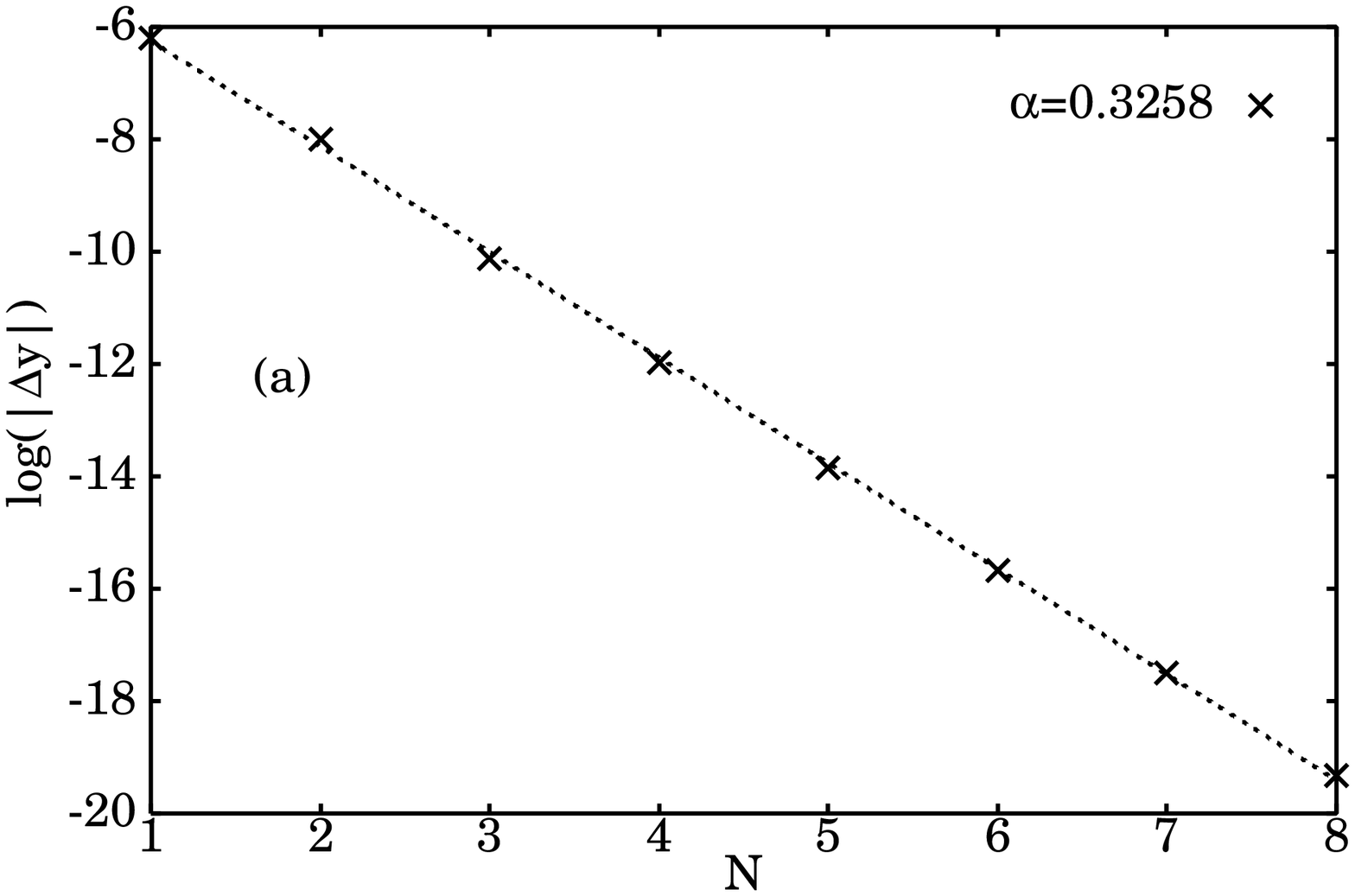}%
\includegraphics[width=8cm]{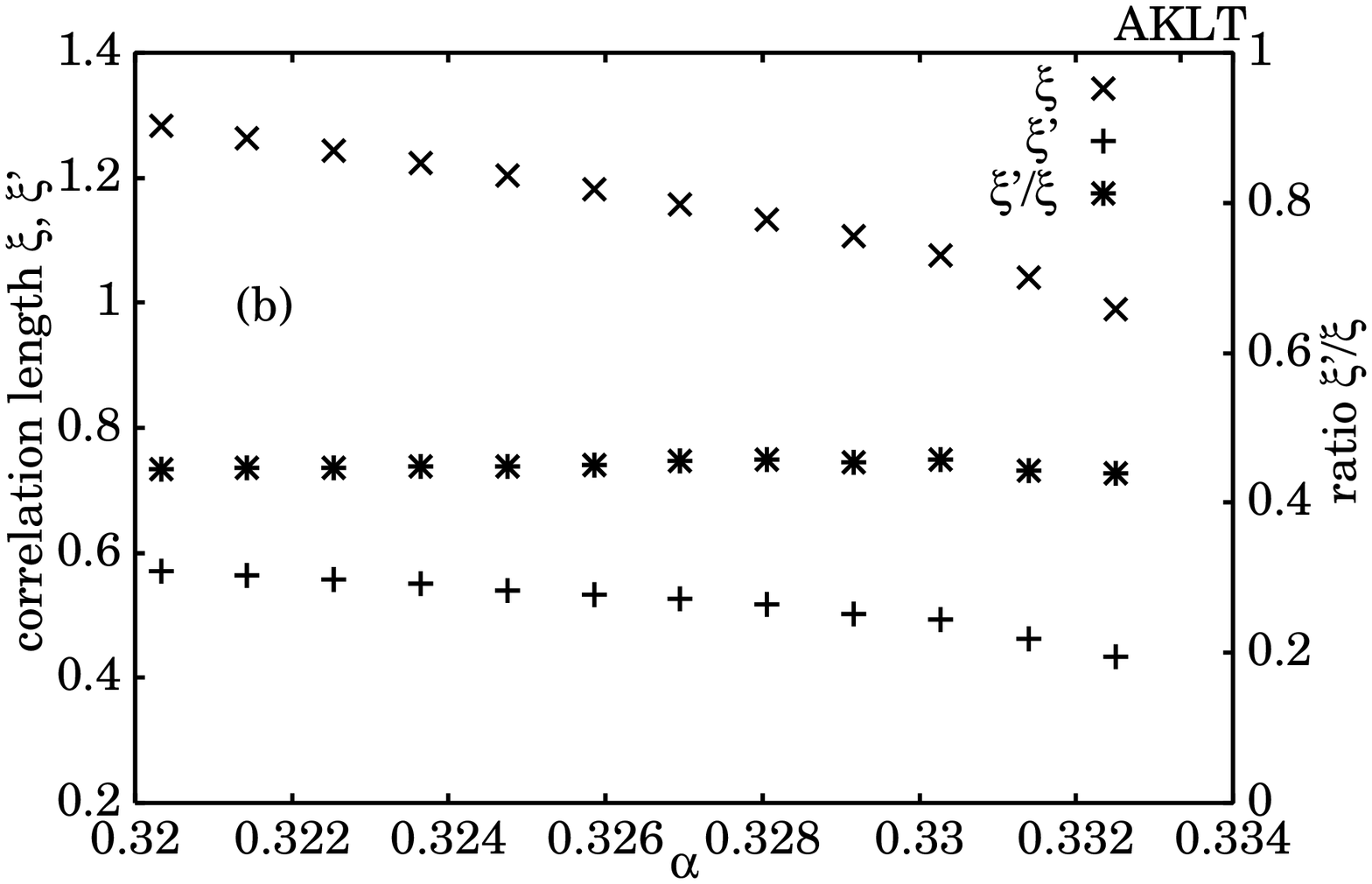}
\end{center}
\caption{\label{fig_dy}(a) The logarithm of $|\Delta y|$ (Eq. \eqref{eq_dy})
are plotted against $N$. The dotted line is drawn for a guide to the eye.
(b) Two different
 correlation lengths
and the ratio $\xi'/\xi$ are plotted in the commensurate region.}
\end{figure}

So far we have excluded data smaller than $N=8$ since these data differ from 
Eq. \eqref{eq_fit}.
Then we proceed to study small size corrections from Eq. \eqref{eq_fit}.
For a rough estimate, we consider the difference
\begin{equation}
\Delta y (N)= \Ets (N)/N - (c'_0 + c'_1/N) \label{eq_dy}
\end{equation}
for small $N$, although we use $c'_0$ and $c'_1$, 
which are determined from the data of $N=14,15$.
Figure \ref{fig_dy} (a) shows $\log (|\Delta y|)$ for $1 \le N \le 8$ 
at $\alpha=0.3258$ in the commensurate region.
Since the logarithm of  $\Delta y$ decreases linearly with $N$,
the difference $\Delta y$ results in 
$\Delta y \sim \exp(-N/\xi')$.
We estimate $\xi'$ as shown in Fig. \ref{fig_dy}
(b),
and then we find
$\xi'\sim \xi/2$ (The correlation length $\xi$ is obtained in the
previous study\cite{MN2006})\cite{memo3}.

In the incommensurate region,
we see the oscillatory behavior again.
We roughly estimate the wave number of this oscillation,
and then we find $q'_{\rm IC} \sim 2(q_{\rm
IC}+\pi)$\cite{memo4}.
We will show a detailed calculation in another paper.

\section{Summary}
We have shown that 
the principal oscillatory behaviors,
which are proportional to $\Sd_{\rm L}\cdot\Sd_{\rm R}$,
among the 
triplet and singlet energy spectra under open boundary
 conditions cancel out
in the commensurate and incommensurate subphases.

We have found that
the energy spectra of singlet and triplet states under open boundary
conditions
consist of the bulk, surface, and edge spin energies: 
\begin{equation}
%\fl
E_{\rm C}/N = B + S/N +
\lk{\rm C}|{\cal H}_{\rm eff}|{\rm C}\>,
\qquad ({\rm C=\{S, T\}})
\end{equation}
where $B$ and $S$
are the bulk and surface energies,
respectively.
%The fact that the energy gap of edge states does not 
%possess such a constant term and a
%linear term proportional to $1/N$
%leads to $B=c_0$ and $S=c_1$.
%The parameters $c_0$ and $c_1$ are model-dependent.

Considering small $N$,
we have found an exponentially decaying correction term,
the correlation length and the
incommensurate wave number
of which differ by factor 2 from those obtained by the energy gap of
edge states.
One possibility is that the correction will be 
$O((\Sd_{\rm L}\cdot\Sd_{\rm R})^2)$.
Therefore, 
we will need to improve the SA theory so as to contain such higher terms.

We observe similar results for the 
$S=1$ next-nearest-neighbor model.
Thus the cancellation of triplet-singlet energies 
can be found in general spin gap systems.

\begin{acknowledgments}
The numerical calculation in this work is based on the program packages
TITPACK version 2, developed by Professor H. Nishimori.

This research is partially supported by a Grant-in-Aid for Scientific
Research (C), 18540376 (2006), from the Ministry of Education, Science,
Sports and Culture of Japan.
\end{acknowledgments}

\bibliography{hfm2006}

\end{document}